 \documentclass[11pt]{article}
 \usepackage[dvips]{graphicx}
 
\def\D{{\cal D}}
\newcommand{\be}{\begin{equation}}
\newcommand{\ee}{\end{equation}}
\newcommand{\bea}{\begin{eqnarray*}}
\newcommand{\eea}{\end{eqnarray*}}
\newcommand{\beqa}{\begin{eqnarray}}
\newcommand{\eeqa}{\end{eqnarray}}
\newcommand{\bref}[1]{(\ref{#1})}
\vspace{1cm}
\begin{document}
\begin{titlepage}
\vspace{4\baselineskip}
\begin{flushright}
MISC-2010-17
\end{flushright}
\vspace{0.6cm}
\begin{center}
{\Large\bf  Toda Molecule and Tomimatsu-Sato Solution}\\
\vspace{5mm}
{\Large\bf$\sim$  Towards the complete proof of Nakamura's conjecture$\sim$}
\end{center}
\vspace{1cm}
\begin{center}
{\large Takeshi Fukuyama$^{a,b,}$
\footnote{E-mail:fukuyama@se.ritsumei.ac.jp}}
and
{\large Kozo Koizumi$^{b,}$
\footnote{E-mail:kkoizumi@cc.kyoto-su.ac.jp}}
\end{center}
\vspace{0.2cm}
\begin{center}
${}^{a}$ {\small \it Department of Physics and R-GIRO, Ritsumeikan University,
Kusatsu, Shiga,525-8577, Japan}\\[.2cm]
${}^{b} $ {\small \it Maskawa Institute for Science and Culture,
Kyoto Sangyo University, Kyoto 603-8555, Japan}
\medskip
\vskip 10mm
\end{center}
\vskip 10mm
\begin{abstract}
We discuss the Nakamura's conjecture stating that the Tomimatsu-Sato black hole solution with integer deformation parameter $n$ is composed of the special solutions of the Toda molecule equation at the $n$-th lattice site.
From the previous work, in which the conjecture was partly analytically proved, we go further towards final full proof by rearranging the rotation parameter.
The proof is explicitly performed for the highest and lowest orders. Though the proof for all orders remains still unsolved,
the prospect to the full proof becomes transparent and workable by our method.

\end{abstract}
\vspace{0.1cm} 

\vskip 0.8cm

\end{titlepage}

\section{Introduction}
There are close and unexpected relations between a variety of integrable systems.
The interplay between black hole and the Toda molecule solutions is one of such examples.
A.~Nakamura found an important relation \cite{anaka} between the Tomimatsu-Sato
 (hereafter we refer it as TS) solutions \cite{T-S} and the Toda molecule solutions \cite{Toda}. That is, he asserts that a series of the TS solutions with integer deformation parameter $n$ and rotation parameters $p,\ q$ ($p^2+q^2=1$) are obtained from the special solutions of the Toda molecule equation at the $n$-th lattice site (Nakamura's conjecture).
The Ernst equation is summarized as two sets of equations, \{Eqs.~(\ref{eq:tsdec1}) and (\ref{eq:tsdec2})\} and \{Eqs.~(\ref{eq:tsdec3}) and
(\ref{eq:tsdec4})\}. Nakamura checked that this conjecture is numerically satisfied for small $n$.
One of the present authors (T.F.) tried to prove analytically this conjecture for general $n$ \cite{F-K-Y}.

In \cite{F-K-Y}, we proved for general $n$ case that the special solutions of Toda molecule satisfy the first set of equations without using the explicit form of the solution. However, the second set was proved to satisfy for the restricted case of $q=0$ by using the explicit form of the solution.
The $q=0$ black hole is corresponding to the (extended)  Weyl solutions. (The solution in the case $q=0,~n=1$ corresponds to the Schwarzschild solution.)
Thus the full proof has not been solved for the generic case, $q\not=0$.
For general $n$, the two solutions of Toda molecule equation, $g_n$ and $f_n$ (see Eq.~\bref{eq:bitmeq1}), are given by the polynomials of $p$ and $q$ of homogeneous degree of $n$ and $n-1$, respectively,
that is, $p^iq^{n-i},\ (i=0,\cdots n)$ and $p^iq^{n-i-1},\ (i=0,\cdots n-1)$.
If we change these parameters $p,\ q$ (independent parameter is one) to $t$ defined by Eq.~\bref{t},
the special solutions of Toda molecule equation $g_n$ and $f_n$ are described by the Laurent polynomials whose highest (lowest) degrees are $n$ ($-n$) and $n-1$ ($-n+1$), respectively. We give the explicit form of the Nakamura's conjecture order by order of $t$. In this paper we prove the second set of equations at the highest and lowest orders of $t$.
This may be the step towards the complete proof.

This paper is organized as follows.
In the next section we briefly review our previous work \cite{F-K-Y}.
Sec.~3 is the central part of this paper, where the Toda molecule equations and the Nakamura's conjecture are expanded by the Laurent polynomials and its conjecture is proved at the highest and lowest orders of $t$. Some detailed calculations are developed in the Appendix.
In Sec.~4, we discuss on the implications of obtained results, future prospect, and the relations with the other approaches. 

\section{Nakamura's conjecture and our previous results}
The two dimensional Toda molecule equation is described by 
\be
\frac{\partial^2}{\partial S\partial T}\log V_n\equiv (\log V_n)_{ST}=V_{n-1}-2V_n+V_{n+1}.
\ee
Here $S$ and $T$ are light cone coordinates related with the Cartesian coordinates, $X$ and $Y$, by
\be
S=\frac{1}{2}(X+Y),~~T=\frac{1}{2}(X-Y)
\ee
and $n$ indicates the $n$-th lattice site.
If we introduce $\tau$ function defined by
\be
V_n=(\log \tau_n)_{ST},
\ee
the Toda molecule equation is expressed in terms of the Hirota's bilinear forms \cite{hiro} as 
\begin{equation}
  D_SD_T \tau_n \cdot \tau_n - 2 \tau_{n + 1} \tau_{n - 1} = 0,
\label{eq:bitmeq1}
\end{equation}
where the Hirota derivative is defined by
\beqa
    &&D_A(f\cdot g)=(\partial_A f)g-f(\partial_A g),\nonumber\\
    &&D_AD_B(f\cdot g)=(\partial_A\partial_B f)g-(\partial_A f)(\partial_B g)
      -(\partial_B f)(\partial_A g)+f(\partial_A\partial_B g)
\eeqa
and so on. The bilinear form is very useful for the integrable system, 
which is also the case in the present problem. $n$ is positive integer and the boundary condition is chosen as $\tau_0 = 1$ corresponding to the finite and semi-infinite lattices.
The general solution of Eq.~(\ref{eq:bitmeq1}) is expressed
in a form of the two-directional Wronskian \footnote{Historically the relation between the Hirota type equation (\ref{eq:bitmeq1}) and the two-directional Wronskian was already found by Darboux \cite{dar}.} \cite{dar,hos}
\begin{equation}
\tau_n = \det \pmatrix{
                   \Psi     & L_- \Psi     & \ldots & L_-^{n - 1} \Psi
\cr                L_+ \Psi & L_+ L_- \Psi & \ldots & L_+ L_-^{n - 1} \Psi
\cr                \vdots   & \vdots       & \vdots & \vdots
\cr        L_+^{n - 1} \Psi & L_+^{n - 1} L_- \Psi & \ldots &
   L_+^{n - 1}L_-^{n - 1} \Psi \cr }
\label{eq:tmsol}
\end{equation}
with the boundary condition $\tau_0 = 1$ and initial condition $\tau_1=\Psi$. 
Here $L_+(L_-) \equiv \partial_S (\partial_T)= \frac{\partial}{\partial X} + \frac{\partial}{\partial Y}(\frac{\partial}{\partial X} - \frac{\partial}{\partial Y})$.
For the finite lattice, the initial condition $\Psi$ takes a form
\be
    \Psi=\sum_{k=1}^{N+1}H_k(S)G_{k}(T).
\ee
So $V_n$ satisfies the boundary condition
\be
V_0=V_{N+1}=0,
\ee
where $N$ stands for the total number of lattice sites. 
In the present paper, the semi-infinite lattice 
corresponding to the $N\to\infty$ case is treated.
We first show how the solution Eq.~\bref{eq:tmsol} of Eq.~\bref{eq:bitmeq1} appears as a result of a Pfaffian identity
for the help of later discussions, though it is a known fact.

We introduce $\D$ as a determinant of $(n+1)\times(n+1)$ matrix $\tau_{n+1}$;
\begin{equation}
  \D ~\equiv~\tau_{n+1} .  \label{eq:D}
\end{equation}
The minor $\D\biggl[{i \atop j}\biggl]$ is defined by deleting the $i$-th row
and the $j$-th column from $\D$. Similarly $\D\biggl[{{i,k} \atop {j,l}}\biggl]$ is defined by deleting
the $i$ and $k$-th rows and the $j$ and $l$-th columns from $\D$ and so on.
The Toda molecule equation (\ref{eq:bitmeq1}) is now expressed as
\begin{equation}
  \D\biggl[{n \atop n}\biggl] \D\biggl[{{n + 1} \atop {n + 1}}\biggl] -
  \D\biggl[{{n + 1} \atop n}\biggl] \D\biggl[{n \atop {n + 1}}\biggl] -
  \D \D\biggl[{{n,n + 1} \atop {n,n + 1}}\biggl] = 0. \label{eq:tmji}
\end{equation}
It holds since it is nothing but the Jacobi's (Sylvester's) formula for
matrix minors, which is one of Pfaffian identities.
Thus the Toda molecule equation has been reduced to Pfaffian identity
in direct method. It is also the case in the Einstein equation as will
be shown in the following.

The Ernst equation for axially symmetric metric of Einstein equation is
\begin{equation}
(\xi \xi^\ast - 1) \nabla^2 \xi - 2 \xi^\ast \nabla \xi \cdot \nabla \xi = 0,
\label{eq:erunsteq}
\end{equation}
where the star superscript denotes the complex conjugate.
It is shown that
Eq.~\bref{eq:erunsteq} is invariant under the global SU(1,1) transformation
\be
\xi'=\frac{\alpha\xi+\beta^*}{\beta\xi+\alpha^*},~~(|\alpha|^2-|\beta|^2\ne 0).
\label{invariance1}
\ee

Setting 
\be
\xi_n = \frac{g_n}{f_n},
\label{Pade}
\ee
the Nakamura's conjecture on the TS solutions consists of two
ingredients \cite{anaka};

\begin{itemize}
\item[(i)] 
Eq.~(\ref{eq:erunsteq}) has a decomposition into two sets \cite{anaka}
\begin{eqnarray}
  & &D_x (g_n \cdot f_n - g_n^\ast \cdot f_n^\ast) = 0, \label{eq:tsdec1}\\
  & &D_y (g_n \cdot f_n + g_n^\ast \cdot f_n^\ast) = 0, \label{eq:tsdec2}
\end{eqnarray}
and
\begin{eqnarray}
  & &F (g_n^\ast \cdot f_n) = 0, \label{eq:tsdec3}\\
  & &F (g_n^\ast \cdot g_n + f_n^\ast \cdot f_n) = 0. \label{eq:tsdec4}
\end{eqnarray}

Here the bi-linear operator $F$ is
\begin{equation}
 F = (x^2 - 1) D_x^2 + 2 x \partial_x + (y^2 - 1) D_y^2 + 2 y \partial_y +
c_n \label{eq:F}
\end{equation}
and  $x$ and $y$ are usual prolate spheroidal coordinates defined by
\be
\partial_X=(x^2-1)\partial_x ~\mbox{and}~ \partial_Y=(y^2-1)\partial_y.
\ee
\medskip

\item[(ii)] From the solutions of the Toda molecule equation $\tau_n$,
a set of TS solutions $\xi_n=\frac{g_n}{f_n}$ with deformation parameter $n$ is obtained as
\begin{equation}
  g_n~ =~\tau_n~=~\D\biggl[{{n + 1} \atop {n + 1}}\biggl], \hspace{1cm}
  f_n~=~\tau_{n-1}\mid_{_{\psi\rightarrow L_+L_-\psi}}~=~
\D\biggl[{{1,{n + 1}} \atop {1,{n + 1}}}\biggl] \label{eq:gnfn}
\end{equation}
by choosing the arbitrary function $\Psi$
and the constant $c_n$ in (6) as
\begin{equation}
\label{eq:choipsi}
  \Psi=\psi \equiv  p x - i q y,~~~~~p^2~+~q^2~=~1,~~~~~c_n~=~-~2~n^2. \label{eq:psi}
\end{equation}
\end{itemize}

The first set Eqs.~(\ref{eq:tsdec1}) and
(\ref{eq:tsdec2}) are proved in the general case whose detailed proof should be referred to \cite{F-K-Y}. 
While, the second set Eqs.~(\ref{eq:tsdec3}) and (\ref{eq:tsdec4}) are shown for a restricted case
of $q=0$ in \cite{F-K-Y}. The present paper is its extension to generic case $q\neq 0$ and we review for $q=0$ first.
\vskip 6mm

To prove the second set Eqs.~(\ref{eq:tsdec3})
and (\ref{eq:tsdec4}), the explicit
form of $\psi$ in Eq.~(\ref{eq:psi}) is required in contrast to the case of the first set.
$g_n(f_n)$ is the determinant of matrix whose $(i,j)
\left((i-1,j-1)\right)$
element is
\begin{eqnarray}
 L_+^{i-1} L_-^{j-1} \psi &=& L_+^{i-1} L_-^{j-1} (p x - i q y) \nonumber
 \\                       &=& p W_{i+j-1}(x) + (-1)^j i q W_{i+j-1}(y),
                          \end{eqnarray}
                          where
                          \begin{equation}
                           W_{n + 1}(z) = (z^2 - 1) \frac{d}{dz} W_{n}(z)
                           \hspace{1cm} {\rm with} \hspace{1cm}
 W_1(z) = z. \label{eq:W}
 \label{eq:gfij}
\end{equation}
The polynomial expression for $W_n(z)$ is given by
\be
    W_{n}(z)=\sum_{m=0}^{n-2}\left\{\sum_{l=0}^{m}(-1)^l(m-l+1)^{n-1}\ {{n}\choose{l}}\right\}(z+1)^{m+1}(z-1)^{n-m-1},\quad(n\geq 2),
\ee
where ${n}\choose{l}$ is the binomial coefficient.

In case of $p=1$ and $q=0$, we have $\psi = x$. Then 
$g_n$ and $f_n$ are real functions depending only on $x$.
Explicit forms of $g_n$ and $f_n$ are
\begin{equation}
  g_n = \det \pmatrix{
                     W_1(x)      & W_2(x)      & \ldots  & W_n(x)        \cr
                   W_2(x)      & W_3(x)      & \ldots  & W_{n+1}(x)    \cr
                   \vdots   & \vdots   & \vdots  & \vdots     \cr
                   W_n(x)      & W_{n+1}(x)  & \ldots  & W_{2n-1}(x)   \cr
                                      },~~~~~
  f_n = \det \pmatrix{
                          W_3(x)      & \ldots  & W_{n+1}(x)    \cr
                          \vdots   & \vdots  & \vdots     \cr
                          W_{n+1}(x)  & \ldots  & W_{2n-1}(x)   \cr
                                      }. \label{eq:gn}
\end{equation}

Using Eq.~(\ref{eq:gfij}) we can evaluate the determinant,
\begin{eqnarray}
  g_n &=& (x^2-1)^{\frac{n(n-1)}{2}}\det \pmatrix{
    x          & 1          & 0       & 0    & \ldots    & 0   \cr
 x^2 - 1       & 2 x        & 2       & 0    & \ldots    & 0   \cr
2 x (x^2 - 1)  & 6 x^2 -2   & 12 x    & 12   & \ldots    & 0   \cr
\vdots         & \vdots & \vdots  & \vdots  & \ldots & \vdots  \cr
                                      } \nonumber\\
&=&  \frac{A_n}{2}(x^2 -1)^{\frac{n (n - 1)}{2}}
\biggl((x + 1)^n + (x - 1)^n\biggr), \label{eq:gnx}
      \end{eqnarray}
where the coefficient $A_n$ is
\begin{equation}
    A_n=\left(\prod_{k=0}^{n-1}\Gamma(n-k)\right )^2=(n-1)^2(n-2)^4\cdots 2^{2(n-1)}.\label{eq:An}
\end{equation}

Eq.~(\ref{eq:gnx}) is proved by induction.
For $n = 1$ and $n=2$ cases Eq.~(\ref{eq:gnx}) holds.
Assuming it for $n = l-1$ and $n=l$ cases,
we prove it for the case of $n = l + 1$.
Applying Jacobi's formula (\ref{eq:tmji}) to $g_n$ we obtain,
corresponding to the Toda molecule equation in Eq.~(\ref{eq:tmsol}),
\begin{equation}
    g_{l - 1}~g_{l + 1} = (L_X^2~ g_l)~g_l - (L_X ~g_l)^2, \label{eq:gjaco}
\end{equation}
where $L_X \equiv (x^2 - 1) \partial_x$. 
Using the assumed forms for $g_l$ and $g_{l-1}$ we find the expected form for $g_{l+1}$;
\begin{equation}
   g_{l + 1} =  \frac{A_{l + 1}}{2} (x^2 -1)^{\frac{(l + 1) l}{2}}
 \biggl((x + 1)^{l + 1} + (x - 1)^{l + 1}\biggr).
\end{equation}
Here we have used an equality obtained from the definition of $A_n$
in Eq.~(\ref{eq:An}):
\begin{equation}
  A_{n - 1} A_{n + 1} = n^2 A_n.
\end{equation}
Thus Eq.~(\ref{eq:gnx}) is proved for $n = l + 1$.
Quite analogously $f_n$ is shown to be
\begin{equation}
 f_n =  \frac{A_n}{2}(x^2 -1)^{\frac{n (n - 1)}{2}}
\biggl((x + 1)^n -  (x - 1)^n\biggr). \label{eq:fnx}
\end{equation}

The second set Eqs.~(\ref{eq:tsdec3}) and (\ref{eq:tsdec4}) are
 \begin{eqnarray}
  & &F (g_n \cdot f_n) = 0, \nonumber \\
   & &F (g_n \cdot g_n + f_n \cdot f_n) = 0, \label{eq:wdec34}
   \end{eqnarray}
and $F$ becomes in $q=0$ and $c_n=-2n^2$ case:
\be
 F a \cdot b = \frac{1}{x^2 - 1} \biggl((L_X^2 a) b + a (L_X^2 b) -
2 (L_X a) (L_X b)\biggr) - 2 n^2 a b.
\label{F}
\ee

It is straight forward to show $g_n$ and $f_n$ in Eqs.~(\ref{eq:gnx}) and
(\ref{eq:fnx}) satisfy Eq.~(\ref{eq:wdec34}).

\vskip 6mm
\section{Proof of Nakamura's Conjecture at the highest and lowest orders}

In the paper \cite{F-K-Y}, Eqs. \bref{eq:tsdec1} and \bref{eq:tsdec2} are fully proved by reducing them to a Pfaffian identity. On the other hand, Eqs. \bref{eq:tsdec3} and \bref{eq:tsdec4} are proved only in the restricted case $p=1,\ q=0$. It is due to the fact that the operator $F$ of Eq. \bref{F} is rather complicated two variable functions and is not described by complete bilinear forms, which makes the use of algebraic relations very difficult.  

In this section we adopt tedious but workable approach to go beyond $p=1,\ q=0$ case, making use of already proved first set of equations.
That is, rewriting $p,\ q$ as Eq. \bref{t} and expanding $f_n,g_n$ as the powers of $t$, we prove the Nakamura's conjecture order by order of $t$.
Unfortunately, the explicit proof is restricted in case of the highest and lowest order of $t$ but it may give the route towards final proof.

We will explain our proof step by step.
In the first step, we observe how the SU(1,1) global symmetry of the Ernst equation reflects into Toda molecule equation. We find a recursion formula mixed of the functions $g_n$ and $f_n$. The parameter $t$ is introduced instead of $p,\ q$ in the second step. The functions $g_n$ and $f_n$ are expressed in the form of the Laurent polynomials of $t$ and the properties of their expansion coefficents are investigated. In the third step, using the expressions for $g_n$ and $f_n$ of the step 2, Toda molecule equations and the recursion formula obtained in the step 1 are represented as the equations in every order of $t$.  In the last step, the Nakamura's conjecture at the highest and lowest orders of $t$ are proved with the polynomial expressions for the highest and lowest terms of $g_n$ and $f_n$.\\

\noindent
\fbox{{\it Step }1}\quad If the Nakamura's conjecture is true, the global symmetry of Eq.~\bref{invariance1} in the Ernst equation must be reflected in the Toda molecule equation.
Eq.~\bref{invariance1} with Eq. \bref{Pade} is rewritten as
\be
\frac{g_n'}{f_n'}=\frac{\alpha g_n+\beta^*f_n}{\beta g_n+\alpha^*f_n}.
\label{tf1}
\ee
This may be divided into the following two transformations:
\begin{eqnarray}
\label{inv2a}
g_n'&=&\alpha g_n+\beta^\ast f_n, \\
f_n'&=&\beta g_n+\alpha^\ast f_n.
\label{inv2b}
\end{eqnarray}
Indeed it is easily shown that two sets of Eqs. \bref{eq:tsdec1}-\bref{eq:tsdec4} in the Nakamura's conjecture remain invariant under the above transformations Eqs.~\bref{inv2a} and \bref{inv2b}. Furthermore, it turns out that the transformed functions $g_n^\prime$ and $f_n^\prime$ are also solutions of the Toda molecule equations, {\it i.e.} any linear superposition of the solutions $g_n$ and $f_n$ given by Eq.~\bref{eq:gnfn} satisfies the Toda molecule equation. The proof for the linear property is confirmed by several ways. As the first proof, we consider that the transformation $\psi\to\psi+\frac{b}{a}$ in Eq.~(\ref{eq:tmsol}) generates different solution of Toda molecule equation. Under the transformation, the function $\tau_n$ becomes $\tau_n+\frac{b}{a}\tau_{n-1}|_{\psi\to L_+L_-\psi}$. 
Taking the transformation and the rescaling of the tau function into account, 
one can show that the function $a\tau_n+b\tau_{n-1}|_{\psi\to L_+L_-\psi}$ satisfies the Toda molecule equation, where $a$ and $b$ are any complex constants. Because $\tau_n=g_n$ and $\tau_{n-1}|_{\psi\to L_+L_-\psi}=f_n$ under the choices Eq.~(\ref{eq:choipsi}), it is shown that $a g_n+b f_n$ is also a solution of Toda molecule equation. As the second proof, it follows from this linear combination Eq.~\bref{inv2a} (Eq.~\bref{inv2b}) that
\be
\label{fgrel}
    D_SD_T f_n\cdot g_n=f_{n+1}g_{n-1}+f_{n-1}g_{n+1},
\ee
which can be reduced to a Pfaffian identity. This equation is useful for getting the recursion formula Eq. \bref{gfhtodeq} and proving the polynomial forms Eqs. \bref{fhterm} and \bref{flterm}.\\

\noindent
\fbox{{\it Step }2}\quad Here let us consider the expansions of the functions $g_n$ and $f_n$ in terms of the parameters $p$ and $q$. For general $n$, $g_n$ and $f_n$ are given by the polynomials of $p$ and $q$ of homoheneous degree of $n$ and $n-1$, respectively.
However, they have rather complicated forms in terms of $p$ and $q$. So let us introduce 
one parameter $t$ defined by
\be
    p=\frac{t+t^{-1}}{2},\quad q=\frac{t-t^{-1}}{2i}
\label{t}.
\ee
Then $\psi$ is rewritten as
\be
    \psi=t\left(\frac{x-y}{2}\right)+t^{-1}\left(\frac{x-y}{2}\right)=tv+t^{-1}u,\quad
    u=\frac{x+y}{2},\ v=\frac{x-y}{2}.
\ee

To show the $t$ dependence of $g_n,f_n$ explicitly, we write them as
\be
    g_n=g_n(x,y;t),\quad f_n=f_n(x,y;t).
\label{(n)}
\ee
Then, $g_n(x,y;t)$ and $f_n(x,y;t)$ are expressed by the Laurent polynomials of $n$ and $n-1$ 
orders on $t$, respectively:
\be
    g_n(x,y;t) = \sum_{m=-n}^{n}{\tilde g}_n^{(m)}(x,y)t^m,\quad
   f_n(x,y;t) = \sum_{m=-n+1}^{n-1}{\tilde f}_n^{(m)}(x,y)t^m.
\ee
Here the functions ${\tilde g}_n^{(m)}(x,y)$ and ${\tilde f}_n^{(m)}(x,y)$ are real functions and have the following properties
\begin{eqnarray}
    &&g_n^\ast= g_n(x,y;t^{-1}),\quad f_n^\ast=f_n(x,y;t^{-1}),
    \label{prop1}\\
    &&g_n^\ast=g_n(x,-y;t),\quad f_n^\ast=f_n(x,-y;t),
    \label{prop2}\\
   &&g_n(x,y;-t)=(-1)^n g_n(x,y;t),\quad f_n(x,y;-t)=(-1)^{n-1} f_n(x,y;t),
   \label{prop3}\\
   &&g_n(x,y;it)=(-i)^{n^2} g_n(y,x;t),\quad f_n(y,x;it)=(-i)^{n^2-1} f_n(y,x;t)
   \label{prop4}
\end{eqnarray}  
as are shown from their definitions. It follows from these identities that
\begin{eqnarray}
     &&\hspace{-10mm}{\tilde g}_n^{(n-2m-1)}(x,y)=0,\ \
       {\tilde f}_n^{(-n+2m+1)}(x,y)={\tilde f}_n^{(n-2m-1)}(x,-y),\ \ (m=0,\cdots, n-1),\\  
     &&\hspace{-10mm}{\tilde f}_n^{(n-2m)}(x,y)=0,\ \
     {\tilde g}_n^{(-n+2m)}(x,y)={\tilde g}_n^{(n-2m)}(x,-y),\ \ (m=0,\cdots, n)
\end{eqnarray}
and we obtain the expansions,
\be
\label{gfmodexp}
    g_n(x,y;t) = \sum_{m=0}^{n}{\tilde g}_n^{(n-2m)}(x,y)t^{n-2m},\quad
    f_n(x,y;t) = \sum_{m=0}^{n-1}{\tilde f}_n^{(n-2m-1)}(x,y)t^{n-2m-1}
\ee
and
\be
\label{gfmodexpstar}
    g_n^\ast(x,y;t) = \sum_{m=0}^{n}{\tilde g}_n^{(-n+2m)}(x,y)t^{n-2m},\quad
    f_n^\ast(x,y;t) = \sum_{m=0}^{n-1}{\tilde f}_n^{(-n+2m+1)}(x,y)t^{n-2m-1}
\ee
with
\be
\label{gfidentity}
    {\tilde g}_n^{(-m)}(x,y)={\tilde g}_n^{(m)}(x,-y),\quad
    {\tilde f}_n^{(-m)}(x,y)={\tilde f}_n^{(m)}(x,-y).
\ee
\\

\noindent
\fbox{{\it Step }3}\quad Because the Toda molecule equations for $f_n$ and $g_n$ and Eq.~(\ref{fgrel}) hold for an arbitrary value of the parameter $t$,
we obtain order by order equations of the Laurent expansions of $f_n$ and $g_n$ (see the Appendix A).
The highest order equations of them, corresponding to the $I=0$ case in the Appendix A, are given by
\be
\label{ghtodeq}
   D_SD_T\tilde g_n^{(n)}(x,y)\cdot \tilde g_{n}^{(n)}(x,y)=2\tilde g_{n+1}^{(n+1)}(x,y)\tilde g_{n-1}^{(n-1)}(x,y) ,
\ee
\be
\label{fhtodeq}
   D_SD_T\tilde f_n^{(n-1)}(x,y)\cdot \tilde f_{n}^{(n-1)}(x,y)=2\tilde f_{n+1}^{(n)}(x,y)\tilde f_{n-1}^{(n-2)}(x,y) ,
\ee
and
\be
\label{gfhtodeq}
    D_SD_T\tilde f_n^{(n-1)}(x,y)\cdot \tilde g_{n}^{(n)}(x,y)=\tilde f_{n+1}^{(n)}(x,y)\tilde g_{n-1}^{(n-1)}(x,y)
    + \tilde f_{n-2}^{(n-1)}(x,y)\tilde g_{n+1}^{(n+1)}(x,y).
\ee
By definitions the highest order terms ${\tilde g}_n^{(n)}(x,y)$ and ${\tilde f}_n^{(n-1)}(x,y)$ are written by
the following determinants
\beqa
    &&{\tilde g}_n^{(n)}(x,y)= 
    \left|\begin{array}{cccc}v&L_- v &\cdots &L_-^{n-1}v\\
                             L_+v&L_+L_- v&\cdots&L_+L_-^{n-1}v\\
                             \vdots&\vdots&\ddots &\vdots\\
                             L_+^{n-1}v&L_+^{n-1}L_-v&\cdots &L_+^{n-1}L_-^{n-1}v
           \end{array}\right|,\\
 \noalign{\vskip 2mm}
    &&{\tilde f}_n^{(n-1)}(x,y)=
    \left|\begin{array}{ccc}
                             L_+L_- v&\cdots&L_+L_-^{n-1}v\\
                             \vdots&\ddots &\vdots\\
                             L_+^{n-1}L_-v&\cdots &L_+^{n-1}L_-^{n-1}v
           \end{array}\right|.
\eeqa
Then, all of these equations (\ref{ghtodeq})-(\ref{gfhtodeq}) reduce to Pfaffian identities. 

In general, these functions $g_n$ and $f_n$ should be expressed
as finite polynomials of the prolate spheroidal coordinates $x,\ y$. 
After some numerical caluculations, the highest term ${\tilde g}_n^{(n)}(x,y)$ and the lowest term ${\tilde g}_n^{(-n)}(x,y)$
have the forms:
\be
\label{ghterm}
    {\tilde g}_n^{(n)}(x,y)= 
    \left|\begin{array}{cccc}v&L_- v &\cdots &L_-^{n-1}v\\
                             L_+v&L_+L_- v&\cdots&L_+L_-^{n-1}v\\
                             \vdots&\vdots&\ddots &\vdots\\
                             L_+^{n-1}v&L_+^{n-1}L_-v&\cdots &L_+^{n-1}L_-^{n-1}v
           \end{array}\right|=2^{n(n-1)}A_n  u^{\frac{n(n-1)}{2}}v^{\frac{n(n+1)}{2}}
\ee
and
\be
\label{glterm}
    {\tilde g}_n^{(-n)}(x,y)= 
    \left|\begin{array}{cccc}u&L_- u &\cdots &L_-^{n-1}u\\
                             L_+u&L_+L_- u&\cdots&L_+L_-^{n-1}u\\
                             \vdots&\vdots&\ddots &\vdots\\
                             L_+^{n-1}u&L_+^{n-1}L_-u&\cdots &L_+^{n-1}L_-^{n-1}u
           \end{array}\right|=2^{n(n-1)}A_n u^{\frac{n(n+1)}{2}}v^{\frac{n(n-1)}{2}},
\ee
where $A_n$ is given by Eq.~\bref{eq:An}. 
Also it turns out that the highest term ${\tilde f}_n^{(n-1)}(x,y)$ 
and the lowest term ${\tilde f}_n^{(-n+1)}(x,y)$ are 
\beqa
\label{fhterm}
    &&\tilde f_n^{(n-1)}(x,y)=\left|\begin{array}{ccc}
                             L_+L_- v&\cdots&L_+L_-^{n-1}v\\
                             \vdots&\ddots &\vdots\\
                             L_+^{n-1}L_-v&\cdots &L_+^{n-1}L_-^{n-1}v
           \end{array}\right|\nonumber\\
    &&=\frac{{\tilde g}_n^{(n)}(x,y)}{\sqrt{\pi}}\times\sum_{m=0}^{n-1}\sum_{l=0}^{m}(-1)^{l-m}
       \frac{\Gamma(\frac{2m+1}{2})\Gamma(\frac{2(n-l)+1}{2})}{\Gamma(\frac{2(m-l)+3}{2})\Gamma(l+1)\Gamma(m-l+1)\Gamma(n-m)}u^{2l}v^{-2m-1}\qquad
\eeqa
and
\beqa
\label{flterm}
    &&\tilde f_n^{(-n+1)}(x,y)=\left|\begin{array}{ccc}
                             L_+L_- u&\cdots&L_+L_-^{n-1}u\\
                             \vdots&\ddots &\vdots\\
                             L_+^{n-1}L_-u&\cdots &L_+^{n-1}L_-^{n-1}u
           \end{array}\right|\nonumber\\
    &&=\frac{{\tilde g}_n^{(-n)}(x,y)}{\sqrt{\pi}}\times\sum_{m=0}^{n-1}\sum_{l=0}^{m}(-1)^{l-m}
       \frac{\Gamma(\frac{2m+1}{2})\Gamma(\frac{2(n-l)+1}{2})}{\Gamma(\frac{2(m-l)+3}{2})\Gamma(l+1)\Gamma(m-l+1)\Gamma(n-m)}u^{-2m-1}v^{2l}.\qquad
\eeqa
In deriving the above two equations, already proved first set of equations, Eqs. \bref{eq:tsdec1} and \bref{eq:tsdec2} are very helpful.
The explicit proof of Eqs.~(\ref{ghterm})-(\ref{flterm}) are given by applying the mathematical induction and using the identities (\ref{gfidentity}) as follows. 

In the $n=1$ and $n=2$ cases these expressions of Eqs.~(\ref{ghterm})-(\ref{flterm}) hold. 
Assuming it for $n=l-1$ and $n=l$ cases, we prove them for the case of $n=l+1$.
Substituting the assumed polynomial forms $\tilde g_{l-1}^{(l-1)}$ and $\tilde g_l^{(l)}$ 
into the Toda molecule equation at the highest order,
\be
     \tilde g_{l+1}^{(l+1)}(x,y)=\frac{1}{2\tilde g_{l-1}^{(l-1)}(x,y) }
     (D_SD_T\tilde g_l^{(l)}(x,y)\cdot \tilde g_{l}^{(l)}(x,y)),
\ee
we obtain the expected polynomial form for $\tilde g_{l+1}^{(l+1)}$.
By using the identities (\ref{gfidentity}), it is easily shown Eq.~(\ref{glterm}) is correct.
Eq.~(\ref{gfhtodeq}) is used instead of Eq.~(\ref{fhtodeq}) to prove the parts $\tilde f_n^{(n-1)}(x,y)$ and $\tilde f_n^{(-n+1)}(x,y)$, with Eqs.~(\ref{ghterm}) and (\ref{glterm}). It is also shown that the expressions for $\tilde f_n^{(n-1)}(x,y)$ and $\tilde f_n^{(-n+1)}(x,y)$ are correct by substituting them into Eq.~(\ref{gfhtodeq}) and using the identities (\ref{gfidentity}). 
Thus these polynomial expressions are correct. The linearity property for $g_n$ and $f_n$ ensures that the expression for $\tilde f_n^{(n-1)}(x,y)$ satisfies Eq.~(\ref{fhtodeq}).\\

\noindent
\fbox{{\it Step }4}\quad Let us prove the Nakamura's conjecture at the highest order of $t$. 
Substituting the Laurent expansions Eqs.~(\ref{gfmodexp}) and (\ref{gfmodexpstar}) 
into the conjecture Eqs.~(\ref{eq:tsdec1})-(\ref{eq:tsdec4}), 
it reduces to the equations in every order of $t$ (see the Appendix B).  The highest orders of $t$ for Eqs.~(\ref{eq:tsdec1})-(\ref{eq:tsdec3}) and Eq.~(\ref{eq:tsdec4}) are $2n-1$ and $2n$, respectively. 
They are described as the following equations at the highest order: 
\be
\label{hNak1}
    D_x(\tilde g_n^{(n)}(x,y)\cdot{\tilde f}_{n}^{(n-1)}(x,y)
           -\tilde g_n^{(-n)}(x,y)\cdot \tilde f_{n}^{(-n+1)}(x,y))=0,
\ee
\be
\label{hNak2}
    D_y(\tilde g_n^{(n)}(x,y)\cdot \tilde f_{n}^{(n-1)}(x,y)
           +\tilde g_n^{(-n)}(x,y)\cdot \tilde f_{n}^{(-n+1)}(x,y))=0,
\ee
and
\be
\label{hNak3}
    F(\tilde g_n^{(-n)}(x,y)\cdot \tilde f_{n}^{(n-1)}(x,y))=0,
\ee
\be
\label{hNak4}
    F(\tilde  g_n^{(-n)}(x,y)\cdot \tilde  g_n^{(n)}(x,y))=0.
\ee

Substituting the explicit forms for $\tilde g^{(n)}_n, \tilde g^{(-n)}_n, \tilde f^{(n-1)}_n, \tilde f^{(-n+1)}_n$ of Eqs. (\ref{ghterm})-(\ref{flterm}) into Eqs. (\ref{hNak1})-(\ref{hNak2}) and (\ref{hNak3})-(\ref{hNak4}), we can easily recognize that the latter sets of equations are satisified. Although the validities of Eqs.~(\ref{hNak1}) and (\ref{hNak2}) are clear due to a Pfaffian identiy for the first set  Eqs.~(\ref{eq:tsdec1}) and (\ref{eq:tsdec2}), they are also reproduced from this approach. 

It is also proved by straightforward calculation that the lowest order equations of the Nakamura's conjecture are right though it is easily recognized by using the property of Eq.~\bref{gfidentity}. 
%

\section{Discussions}

In this paper we have discussed a proof of the Nakamura's conjecture on the
TS solutions. 
In the previous work \cite{F-K-Y}, the first set Eqs.~\bref{eq:tsdec1} and \bref{eq:tsdec2} were proved completely without using the explicit forms of $f_n$ and $g_n$.
Whereas, the proof of the second set Eqs.~\bref{eq:tsdec3} and \bref{eq:tsdec4} needed the explicit forms of $f_n$ and $g_n$ and was given for the restricted case $p=1,\ q=0$, deformed but non-rotating black hole, extended Weyl solution. In that case, the key point was that the functions $g_n$ and $f_n$ depends only on $x$ and their explicit polynomial expressions are required. 
This work is the extension of that work.
That is, we have discussed the deformed and rotating black hole ($q\ne 0$) in this paper, extending the previous results to more generic case of $pq\neq 0$. 
To prove the deformed and rotating case we have rearranged the original parameters $p,q$ ($p^2+q^2=1$) to $t$.  Thanks to this arrangement, some properties of the functions $g_n$ and $f_n$ become transparent and workable (for instance Eqs. \bref{prop1}-\bref{prop4}). 
Using the property \bref{prop4}, the explicit forms of functions $g_n$ and $f_n$ are obtained for $p=0,\ q=1$ ($t=\sqrt{-1}$) case and proved to satisfy the Nakamura's conjecture. For this case it is known that the metric indicates extremal TS solution, not asymptotically flat and describes a region of the TS metric near its ergosphere \cite{KK74}.  

For the most generic case, $pq\not=0$, the functions $g_n$ and $f_n$ are embedded in two dimensions and the situation is drastically changed compared with the $pq=0$ case. 
The introduction of $t$ enables us to prove the conjecture order by order of $t$.
In fact, the Laurent expansions of the functions $g_n$ and $f_n$ leads us to the polynomial expressions of $x,\ y$ in Eqs. \bref{ghterm}-\bref{flterm} for the highest and lowest orders of $g_n$ and $f_n$. Consequently, it is proved that the highest and lowest orders of the Nakamura's conjecture are valid by the straightforward calculations. Unfortunately the proof of the conjecture has not been completed at all orders of $t$ though their explicit forms are described in the Appendix.

Indeed, it is possible to recursively determine all order terms of $g_n$ and $f_n$ 
for general $n$ by using Eqs.~\bref{TD1}-\bref{TD9} though it is considerably complex. 
The next highest order equations are described in terms of the highest and next highest orders of $f_n$ and $g_n$ ($I=1$ case in the Appendices A and B).
 Substituting the polynomial forms of Eqs. (56)-(59) into these equations, we have the implicit relations of the next highest functions $g_n^{(n-2)}$ and $f_n^{(n-3)}$. To obtain the explicit forms of them we use Eqs. (B1)-(B6) with $I=1$.

 Furthermore, we observe that the expression for the highest order 
of $f_n$ is connected with the Gamma functions. It is strongly expected that the polynomial expressions 
for $g_n$ and $f_n$ are described by two variables hypergeometric functions.

It is also worth noting that the SU(1,1) symmetry plays 
an important role for the proof of the polynomial expressions Eqs.~\bref{fhterm} and \bref{flterm}. 
The Nakamura's conjecture is invariant under this transformation. 
The physical meanings of its symmetry are that a NUT parameter is generated by the transformation, (for instance the Schwartzschild geometry becomes Taub-NUT geometry by the transformation.). Therefore, if the Nakamura's conjecture is correct, its conjecture is also correct even for a generalization of the TS solutions with a NUT parameter. 

Here some comments are in order on the relation between this theory and the other approaches, especially those using the B\"acklund transformations. Indeed, the B\"{a}cklund transformation has played an important role to generate
infinitely many solutions from an initial solution. There exist the well known the B\"{a}cklund transformations in the stationary axi-symmetric Einstein theory,  so-called discrete
Neugebauer-Kramer mapping \cite{NeuKra} and Ehlers transformations
\cite{Ehl}. These transformations generate the
Neugebauer-Kramer (NK) solution describing the nonlinear superposition
of collinear $N(=1,2,3,\cdots)$ Kerr solitons, which is explicitly
constructed for small $N$ \cite{NKsol}. The wholly analytic proof \cite{VeDa} of the
NK solutions for general $N$ is given in terms of two series of the Y.
Nakamura solutions \cite{YNa} called intermediate solutions (unphysical
solutions). Then the B\"{a}cklund transformations appeared in the NK
solutions give the explicit link between the Y. Nakamura solutions (See
\cite{VeDa} for the details). On the other hand, it is expected that the
TS solutions are derived by taking a limit of NK solutions \cite{NKsol,Tom} (TSNK conjecture), whose proof is left unsolved for the general 2$N$-soliton case. Although one of the 
approaches to the TS solutions as the natural limit of the
NK solutions is also discussed by Dale \cite{Dale04}, it
is not proved. 
Also since its limitation changes the determinant form of
NK solutions (indefinite form), the explicit connections between the
TS solutions in the context of the B\"{a}cklund
transformations are not obtained. In contrast, the Nakamura's
conjecture provides the generation mechanism of the TS solutions in terms of
the Toda molecule equation.
If one can explain the meanings of the generation mechanism by the Toda
molecule equation as the B\"{a}cklund transforamtion, 
then both of Nakamura's conjecture and TSNK
conjecture will be solved. The detail of these connections will be considered elsewhere.

The above mentioned relations are schematically written in Figure 1.

\begin{figure}[htbp]
 \begin{center}
  \includegraphics[width=140mm]{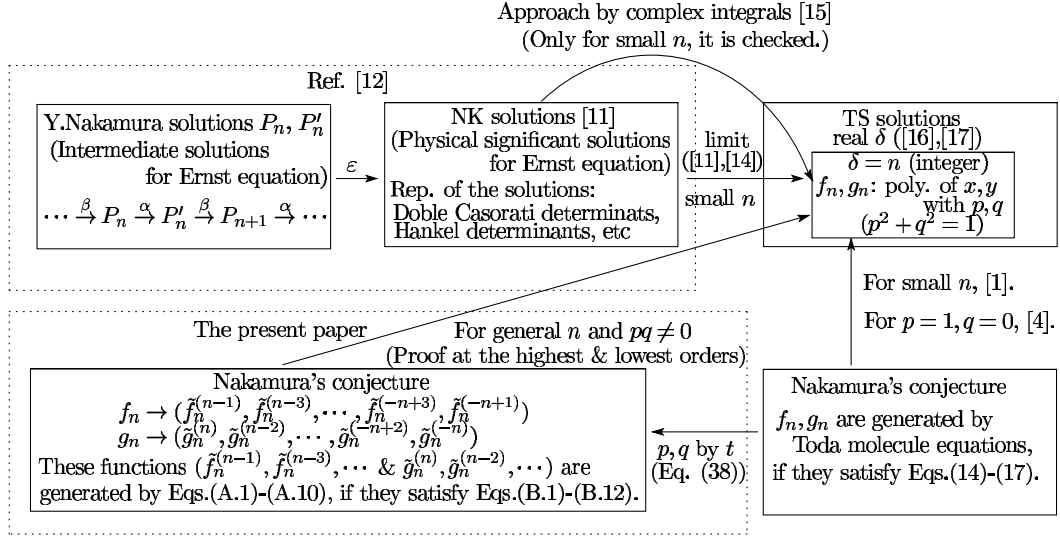}
 \end{center}
 \vspace{-5mm}
 \label{fig1}
 \caption{The relations between ours and the other approaches.
$\alpha,\ \beta,\ \varepsilon$ in the upper left frame indicate the B\"acklund transfomations. $\delta$ in the upper right is a deformation parameter.}
\end{figure}

Thus the study of relations between the Toda molecule equation and the Ernst equation is very important. It explains the reason why the deformation parameter $n$ is limitted as integers for the TS solutions, though generalized Tomimatsu-Sato solutions with non integer deformation parameters were considered by Hori \cite{Horigene} and Cosgrove \cite{Cosgene}. On the physical natures of the geometries of TS solution, for instance, geometrical differences in event horizons and causal structures of odd and even deformation parameters \cite{KodHik} will be also clarified from this approach \cite{Ohta}.

\subsection*{Acknowledgements}
We are grateful to K. Takasaki for useful discussions.
We would like to thank J.H.H. Perk for valuable comments and communications. 
The work of T.F. is supported in part by the Grant-in-Aid for Scientific Research from the Ministry of Education, Science and Culture of Japan (N0.020540282 and No. 21104004). K.K. thanks MISC for financial support and hospitality.

\addcontentsline{toc}{subsection}{Appendix}
\section*{Appendix}

We have seen the functions $g_n$ and $f_n$ are expanded by the one parameter $t$.
In the appendix, we extract order by order equations for the Toda molecule equations and the Nakamura's conjecture by substituting the expressions Eq.~\bref{gfmodexp} into them.

\subsection*{A.\ Order by order equations for Toda molecule equation}
\renewcommand{\theequation}{A.\arabic{equation}}
\setcounter{equation}{0}

The Toda molecule equation for $g_n$ is expressed by
\begin{eqnarray}
    &&\hspace{-10mm}
    D_S D_T\left(\sum_{m=0}^{n}{\tilde g}_n^{(n-2m)}(x,y)t^{n-2m}\right)\cdot
    \left(\sum_{l=0}^{n}{\tilde g}_n^{(n-2l)}(x,y)t^{n-2l}\right)\nonumber\\
    &&\hspace{10mm}=2\left(\sum_{m=0}^{n+1}{\tilde g}_{n+1}^{(n-2m+1)}(x,y)t^{n-2m+1}\right)
    \left(\sum_{l=0}^{n-1}{\tilde g}_{n-1}^{(n-2l-1)}(x,y)t^{n-2l-1}\right)
\end{eqnarray}
in terms of the Laurent expansions by $t$. The parameter $t$ is an arbitrary constant, so that the above equation must hold at every order of $t$.
For the orders $t^{2n-2I},\ (I=0,\cdots 2n)$, we derive the following coupled equations: 
\beqa
&&\hspace{-1cm}
({\rm I})\quad {\rm for}\quad  0\leq I\leq n-1,\nonumber\\
    \noalign{\vskip 2mm}
    &&\hspace{0cm}\sum_{J=0}^{I}
            D_S D_T\ {\tilde g}_n^{(n-2J)}(x,y)\cdot {\tilde g}_n^{(n-2I+2J)}(x,y)\nonumber\\
    \noalign{\vskip -4mm}
      &&\hspace{4cm}=2\sum_{J=0}^{I}
      {\tilde g}_{n+1}^{(n-2J+1)}(x,y){\tilde g}_{n-1}^{(n-2I+2J-1)}(x,y),\label{TD1}\\
        \noalign{\vskip 5mm}
&&\hspace{-1cm}
({\rm I\!I})\quad {\rm for}\quad  I=n,\nonumber\\
 \noalign{\vskip 2mm}
    &&\hspace{0cm}\sum_{J=0}^{n}
            D_S D_T\ {\tilde g}_n^{(n-2J)}(x,y)\cdot {\tilde g}_n^{(-n+2J)}(x,y)\nonumber\\
    \noalign{\vskip -4mm}
      &&\hspace{4cm}=2\sum_{J=1}^{n}
      {\tilde g}_{n+1}^{(n-2J+1)}(x,y){\tilde g}_{n-1}^{(-n+2J-1)}(x,y),\label{TD2}\\
       \noalign{\vskip 5mm}
&&\hspace{-1cm}
({\rm I\!I\!I})\quad {\rm for}\quad  n+1\leq I\leq 2n,\quad y \to -y\quad {\rm in}\  ({\rm I})\quad {\rm and}\quad ({\rm I\!I}).  \label{TD3}  
\eeqa
In the similar way, the Toda molecule equation for $f_n$ becomes the following equations at the order $t^{2n-2I-2},\ (I=0,\cdots,2(n-1))$:
\beqa
&&\hspace{-1cm}
({\rm I\!V})\quad {\rm for}\quad  0\leq I\leq n-2,\nonumber\\
    \noalign{\vskip 2mm}
    &&\hspace{0cm}\sum_{J=0}^{I}
            D_S D_T\ {\tilde f}_n^{(n-2J-1)}(x,y)\cdot {\tilde f}_n^{(n-2I+2J-1)}(x,y)\nonumber\\
    \noalign{\vskip -4mm}
      &&\hspace{4cm}=2\sum_{J=0}^{I}
      {\tilde f}_{n+1}^{(n-2J)}(x,y){\tilde f}_{n-1}^{(n-2I+2J-2)}(x,y),\label{TD4}\\
        \noalign{\vskip 5mm}
&&\hspace{-1cm}
({\rm V})\quad {\rm for}\quad  I=n-1,\nonumber\\
   \noalign{\vskip 2mm}
    &&\hspace{0cm}\sum_{J=0}^{n-1}
            D_S D_T\ {\tilde f}_n^{(n-2J-1)}(x,y)\cdot {\tilde f}_n^{(-n+2J+1)}(x,y)\nonumber\\
    \noalign{\vskip -4mm}
      &&\hspace{4cm}=2\sum_{J=1}^{n-1}
      {\tilde f}_{n+1}^{(n-2J)}(x,y){\tilde f}_{n-1}^{(-n+2J)}(x,y),\label{TD5}\\
       \noalign{\vskip 5mm}
&&\hspace{-1cm}
({\rm V\!I})\quad {\rm for}\quad  n\leq I\leq 2(n-1),\quad y \to -y\quad {\rm in}\  ({\rm I\!V})\quad {\rm and}\quad ({\rm V}).\label{TD6}    
\eeqa
For $t^{2n-2I-1},\ (I=0,\cdots,2n-1)$, Eq.~\bref{fgrel} is reduced to 
\beqa
&&\hspace{-1cm}
({\rm V\!I\!I})\ {\rm for}\quad  0\leq I\leq n-2,\nonumber\\
    \noalign{\vskip 2mm}
    &&\hspace{0cm}\sum_{J=0}^{I}
           D_S D_T\ {\tilde f}_n^{(n-2J-1)}(x,y)\cdot {\tilde g}_n^{(n-2I+2J)}(x,y)\nonumber\\
      &&\hspace{1cm}=2\sum_{J=0}^{I}
      \left({\tilde f}_{n+1}^{(n-2J)}(x,y){\tilde g}_{n-1}^{(n-2I+2J-1)}(x,y)\right.\nonumber\\
      \noalign{\vskip-5mm}
      &&\left.\hspace{4.5cm}+{\tilde f}_{n-1}^{(n-2J-2)}(x,y){\tilde g}_{n+1}^{(n-2I+2J+1)}(x,y)
\right),\label{TD7}\\
        \noalign{\vskip 5mm}
&&\hspace{-1cm}
({\rm V\!I\!I\!I})\ {\rm for}\quad  I=n-1,\nonumber\\
 \noalign{\vskip 2mm}
    &&\hspace{0cm}\sum_{J=0}^{n-1}
           D_S D_T\ {\tilde f}_n^{(n-2J-1)}(x,y)\cdot {\tilde g}_n^{(-n+2J+2)}(x,y)\nonumber\\
      &&\hspace{1cm}=2\sum_{J=0}^{n-1}
      {\tilde f}_{n+1}^{(n-2J)}(x,y){\tilde g}_{n-1}^{(-n+2J+1)}(x,y)\nonumber\\
      \noalign{\vskip-5mm}
      &&\hspace{4.5cm}+\sum_{J=0}^{n-2}{\tilde f}_{n-1}^{(n-2J-2)}(x,y){\tilde g}_{n+1}^{(-n+2J+3)}(x,y),\label{TD8}\\ 
      \noalign{\vskip2mm} 
&&\hspace{-1cm}
({\rm IX})\ {\rm for}\quad  n\leq I\leq 2n-1,\quad y \to -y\quad {\rm in}\  ({\rm V\!I\!I})\quad {\rm and}\quad ({\rm V\!I\!I\!I}). \label{TD9}   
\eeqa
The highest order equations correspond to the $I=0$ case.

\subsection*{B.\ Order by order equations for Nakamura's conjecture}
\renewcommand{\theequation}{B.\arabic{equation}}
\setcounter{equation}{0}

Eqs.~\bref{eq:tsdec1}-\bref{eq:tsdec4} of the Nakamura's conjecture are also rewritten by the Laurent expansions of $t$.
Eq.~\bref{eq:tsdec1} becomes the following equations for the order $t^{2n-2I-1}, (I=0,\cdots,2n-1)$:
\beqa
&&\hspace{-1cm}
({\rm I})\ {\rm for}\quad  0\leq I\leq n-1,\nonumber\\
    \noalign{\vskip 2mm}
    &&\hspace{0cm}\sum_{J=0}^{I}
           D_x\left({\tilde g}_n^{(n-2J)}(x,y)\cdot {\tilde f}_n^{(n-2I+2J-1)}(x,y)\right.\nonumber\\
    &&\hspace{3cm}\left.-{\tilde g}_n^{(-n+2J)}(x,y)\cdot {\tilde f}_n^{(-n+2I-2J+1)}(x,y)\right)=0\label{1stNak11}\\
        \noalign{\vskip 5mm}
&&\hspace{-1cm}
({\rm I\!I})\ {\rm for}\quad  I=n,\nonumber\\
 \noalign{\vskip 2mm}
    &&\hspace{0cm}\sum_{J=1}^{n}
           D_x\left({\tilde g}_n^{(n-2J)}(x,y)\cdot {\tilde f}_n^{(-n+2J-1)}(x,y)\right.\nonumber\\
    &&\hspace{3cm}\left.-{\tilde g}_n^{(-n+2J)}(x,y)\cdot {\tilde f}_n^{(n-2J+1)}(x,y)\right)=0\label{1stNak12}\\
      \noalign{\vskip2mm} 
&&\hspace{-1cm}
({\rm I\!I\!I})\ {\rm for}\quad  n+1\leq I\leq 2n-1,\quad y \to -y\quad {\rm in}\  ({\rm I})\quad {\rm and}\quad ({\rm I\!I})\label{1stNak13}.    
\eeqa
Eq.~\bref{eq:tsdec2} reduces to the following equations at the order $t^{2n-2I-1}, (I=0,\cdots,2n-1)$:
\beqa
&&\hspace{-1cm}
({\rm I\!V})\ {\rm for}\quad  0\leq I\leq n-1,\nonumber\\
    \noalign{\vskip 2mm}
    &&\hspace{0cm}\sum_{J=0}^{I}
           D_y\left({\tilde g}_n^{(n-2J)}(x,y)\cdot {\tilde f}_n^{(n-2I+2J-1)}(x,y)\right.\nonumber\\
    &&\hspace{3cm}\left.+{\tilde g}_n^{(-n+2J)}(x,y)\cdot {\tilde f}_n^{(-n+2I-2J+1)}(x,y)\right)=0\label{1stNak21}\\
        \noalign{\vskip 5mm}
&&\hspace{-1cm}
({\rm V})\ {\rm for}\quad  I=n,\nonumber\\
 \noalign{\vskip 2mm}
    &&\hspace{0cm}\sum_{J=1}^{n}
           D_y\left({\tilde g}_n^{(n-2J)}(x,y)\cdot {\tilde f}_n^{(-n+2J-1)}(x,y)\right.\nonumber\\
    &&\hspace{3cm}\left.+{\tilde g}_n^{(-n+2J)}(x,y)\cdot {\tilde f}_n^{(n-2J+1)}(x,y)\right)=0\label{1stNak22}\\
      \noalign{\vskip2mm} 
&&\hspace{-1cm}
({\rm V\!I})\ {\rm for}\quad  n+1\leq I\leq 2n-1,\quad y \to -y\quad {\rm in}\  (
{\rm I\!V})\quad {\rm and}\quad ({\rm V}).\label{1stNak23}    
\eeqa
For Eq.~\bref{eq:tsdec3} at the order $t^{2n-2I-1}, (I=0,\cdots,2n-1)$, it is
\beqa
&&\hspace{-1cm}
({\rm V\!I\!I})\ {\rm for}\quad  0\leq I\leq n-1,\nonumber\\
    \noalign{\vskip 2mm}
    &&\hspace{0cm}\sum_{J=0}^{I}
           F\ {\tilde g}_n^{(-n+2J)}(x,y)\cdot {\tilde f}_n^{(n-2I+2J-1)}(x,y)=0\label{2stNak11}\\
        \noalign{\vskip 5mm}
&&\hspace{-1cm}
({\rm V\!I\!I\!I})\ {\rm for}\quad  I=n,\nonumber\\
 \noalign{\vskip 2mm}
    &&\hspace{0cm}\sum_{J=1}^{n}
          F({\tilde g}_n^{(-n+2J)}(x,y)\cdot {\tilde f}_n^{(-n+2J-1)}(x,y)=0\label{2stNak12}\\
      \noalign{\vskip2mm} 
&&\hspace{-1cm}
({\rm I\!X})\ {\rm for}\quad  n+1\leq I\leq 2n-1,\quad y \to -y\quad {\rm in}\  ({\rm V\!I\!I})\quad {\rm and}\quad ({\rm V\!I\!I\!I}).\label{2stNak13}    
\eeqa
For Eq.~\bref{eq:tsdec4} at the order $t^{2n-2I}, (I=0,\cdots,2n)$, it is
\beqa
&&\hspace{-1cm}
({\rm X})\ {\rm for}\quad  1\leq I\leq n,\nonumber\\
    \noalign{\vskip 2mm}
    &&\hspace{0cm}\sum_{J=0}^{I}
           F\left({\tilde g}_n^{(-n+2J)}(x,y)\cdot {\tilde g}_n^{(n-2I+2J)}(x,y)\right.\nonumber\\
    &&\hspace{3cm}\left.+{\tilde f}_n^{(-n+2J+1)}(x,y)\cdot {\tilde f}_n^{(n-2I+2J+1)}(x,y)\right)=0\label{2stNak21}\\
        \noalign{\vskip 5mm}
&&\hspace{-1cm}
({\rm X\!I})\ {\rm for}\quad  I=0,\nonumber\\
 \noalign{\vskip 2mm}
    &&\hspace{0cm}
          F({\tilde g}_n^{(-n)}(x,y)\cdot {\tilde g}_n^{(n)}(x,y))=0\label{2stNak22}\\
      \noalign{\vskip2mm} 
&&\hspace{-1cm}
({\rm X\!I\!I})\ {\rm for}\quad  n+1\leq I\leq 2n,\quad y \to -y\quad {\rm in}\  ({\rm X})\quad {\rm and}\quad ({\rm X\!I}).\label{2stNak23}    
\eeqa
Eqs. \bref{1stNak11}-\bref{1stNak23} are corresponding to the first set Eqs.~(\ref{eq:tsdec1}) and (\ref{eq:tsdec2}) of the Nakamura's conjecture. 
As already mentioned, Eqs.~(\ref{eq:tsdec1}) and (\ref{eq:tsdec2}) are proved by using a Pfaffian identity \cite{F-K-Y}. Thus Eqs.~\bref{1stNak11}-\bref{1stNak23} are correct. The highest order equations of the Nakamura's conjecture also correspond to the $I=0$ case.

\end{document}